\documentclass[aps,pra,twocolumn]{revtex4-2}
\usepackage{amsfonts}
\usepackage{amsmath}
\usepackage{mathrsfs}
\usepackage{amssymb}
\usepackage{graphicx}
\usepackage{bm}
\usepackage{color}

\usepackage[colorlinks=true,linkcolor=blue,anchorcolor=blue,citecolor=blue,urlcolor=blue]{hyperref}

\begin{document}
	
\title{Spin-orbit-coupled spin-1 Bose-Einstein condensates in a toroidal trap: even-petal-number necklacelike state and persistent flow}
	
	\author{Keyan Liu}
	\affiliation{International Center of Quantum Artificial Intelligence for Science and Technology (QuArtist) and Department of Physics, Shanghai University, Shanghai 200444, China}
	
	\author{Huaxin He}
	\affiliation{International Center of Quantum Artificial Intelligence for Science and Technology (QuArtist) and Department of Physics, Shanghai University, Shanghai 200444, China}
	
	\author{Chenhui Wang}
	\affiliation{International Center of Quantum Artificial Intelligence for Science and Technology (QuArtist) and Department of Physics, Shanghai University, Shanghai 200444, China}
	
	\author{Yuanyuan Chen}
	\affiliation{International Center of Quantum Artificial Intelligence for Science and Technology (QuArtist) and Department of Physics, Shanghai University, Shanghai 200444, China}
	
	\author{Yongping Zhang}
	\email{yongping11@t.shu.edu.cn}
	\affiliation{International Center of Quantum Artificial Intelligence for Science and Technology (QuArtist) and Department of Physics, Shanghai University, Shanghai 200444, China}

	\begin{abstract}
	Spin-orbit coupling has novel spin-flip symmetries, a spin-1 spinor Bose-Einstein condensate owns meaningful interactions, and a toroidal trap is topologically nontrivial. We incorporate the three together and study the ground-state phase diagram in a Rashba spin-orbit-coupled spin-1 Bose-Einstein condensate with a toroidal trap. The spin-flip symmetries give rise to two different interesting phases: persistent flows with a unit phase winding difference between three components, and necklace states with even petal-number. The existing parameter regimes and properties of these phases are characterized by two-dimension numerical calculations and an azimuthal analytical one-dimension model.
\end{abstract}

\maketitle

\section{Introduction}

In atomic Bose-Einstein condensates (BECs), confinements play a key role in variously pertinent physics.  One of the salient confining geometries is toroidal trap. Considering macroscopic quantum property of  BECs, periodic boundary imposed by toroidal trap naturally gives rise to atomic persistent flows~\cite{Phillips2007}.   Since preparations of ring-shaped trap can be implemented precisely in experiments~\cite{Phillips2007, Campbell2011},  BECs in a toroidal trap become a prototypical system to investigate superfluidity~\cite{Berloff2009, Salerno2011,Baym2013,Campbell2017,Minguzzi2019,Danshita2019}. Furthermore, such confinement can be easily equipped with rotation created by rotating repulsive perturbation~\cite{Campbell2014}. The response of superfluids to rotation in a toroidal trap has been widely investigated~\cite{Mason2010,Perrin2012,Campbell2014,Busch2016,Kavoulakis2017,Brand2019,Cheon2020,Edwards2020,David2021}.

The generalization of  single-component toroidal BECs to multicomponents also draws much attention~\cite{Hadzibabic2013}. The population imbalance and fixed phase relation between multiple components bring persistent flows novel stability features~\cite{Hadzibabic2013,Reimann2009,Reimann2010, Zaremba2013,Recati2014}. It has been found that rich phase diagram and interesting collective excitations can exist in interacting two-component ~\cite{Saito2010,Kavoulakis2015,Malomed2019, Gulliksson2018} and three-component ~\cite{Lundh2013,Weyrauch2013,Kunimi2014} toroidal BECs.

Each component behaves as a pseudo-spin state, therefore,  two components correspond to spin-1/2 and three components can be explained as spin-1.  Pseudo-spins can be arranged to couple with external momentum, which leads to so called spin-orbit coupling. It must be introduced into multi-component BECs artificially~\cite{Spielman2011PRA}.  Experimental realizations of spin-orbit-coupled BECs represent current advances in ultracold atomic physics~\cite{Spielman2011,Pan2015,Engels2015}. The striking feature of spin-orbit-coupled BECs is the spontaneous emergence of striped density patterns as ground states~\cite{Zhai2011,Ho2011,Hu2012, Zhang2012,Stringari2012,Ketterle2017}.   They are due to Bosons condense simultaneously into multiple energy-minimum states,  as distinct from conventional condensations that condense into only one energy minimum.  Putting spin-orbit-coupled BECs in a toroidal trap quickly attracts interests~\cite{Zulicke2012,Reimann2016,Hiroki2017,Busch2017,Yang2017,Li2017,Yang2018}.  Phase diagram of spin-orbit-coupled spin-1/2 BECs in a tight toroidal trap has been identified~\cite{Reimann2016,Hiroki2017,Busch2017}. The non-trivial topology of the toroidal trap generates new features to striped patterns for Raman-induced spin-orbit coupling~\cite{Reimann2016}. While Rashba spin-orbit coupling has same rotating symmetry as the toroidal trap, density modulation for Rashba case is patterned along azimuthal direction  appearing as a necklace~\cite{Hiroki2017,Busch2017}.  The most interesting is such a necklacelike state has an odd number of petals~\cite{Busch2017}. It is also revealed that in Rashba spin-1/2 BECs both two components can support persistent flows with a unit winding number difference between them~\cite{Busch2017}.  Spin-orbit-coupled spin-1 BECs support more enriching phases. The effort in  existing studies on spin-1 has been put into investigating the interplay between spin-orbit coupling and rotation~\cite{Yang2017, Li2017, Yang2018}.  The characteristics of phase diagram for spin-orbit-coupled spin-1 BECs with a toroidal confinement are still lacking. Considering  the existence of the odd-petal-number necklace state in spin-1/2 system, it is natural to ask whether the necklace state still has specific petal-number and what the persistent flow is in spin-1 BECs.

\begin{figure*}[t]
	\includegraphics[width=0.24\textwidth,trim={110 40mm 30mm 80}, clip]{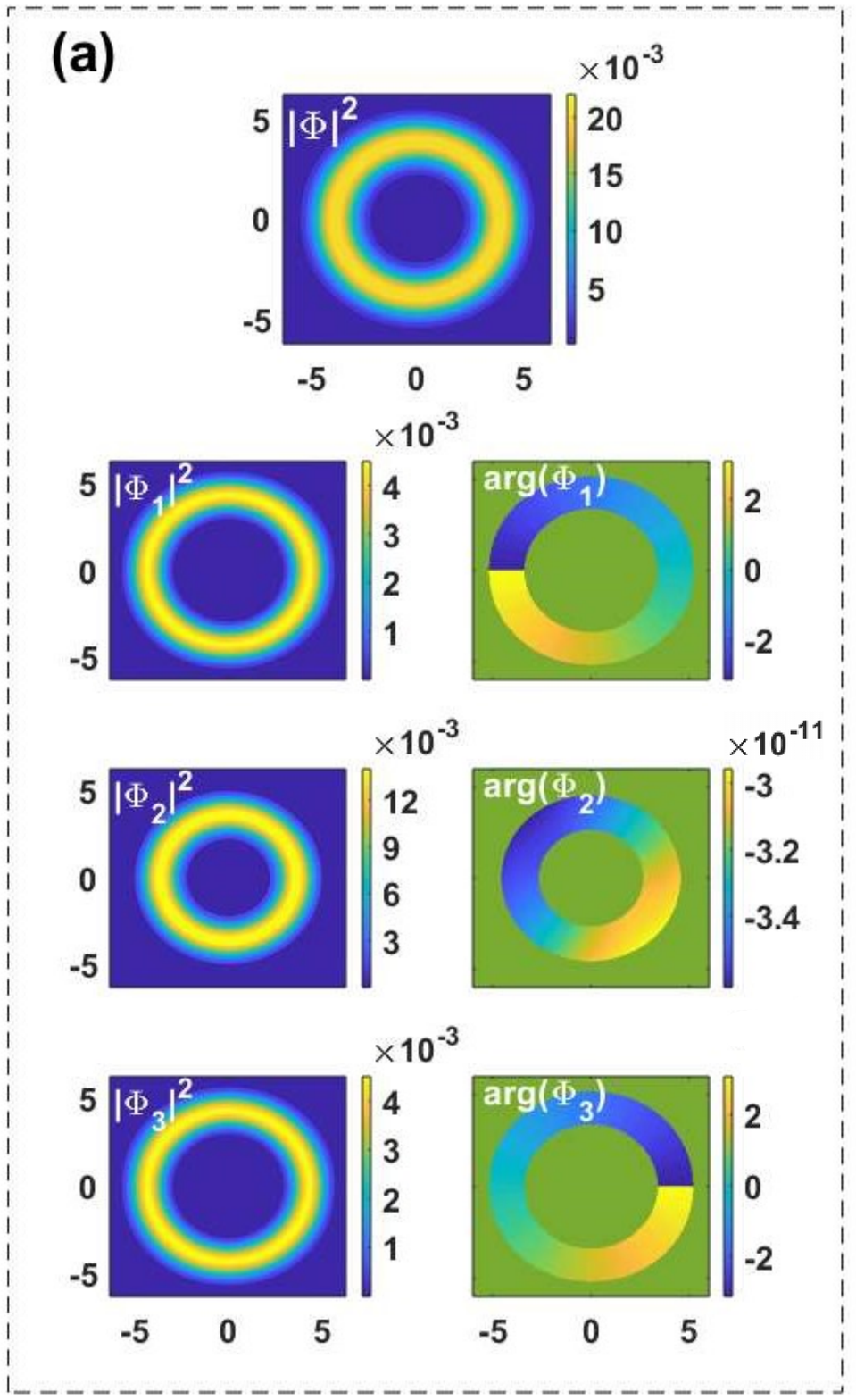}
	\includegraphics[width=0.24\textwidth,trim={110 40mm 30mm 80}, clip]{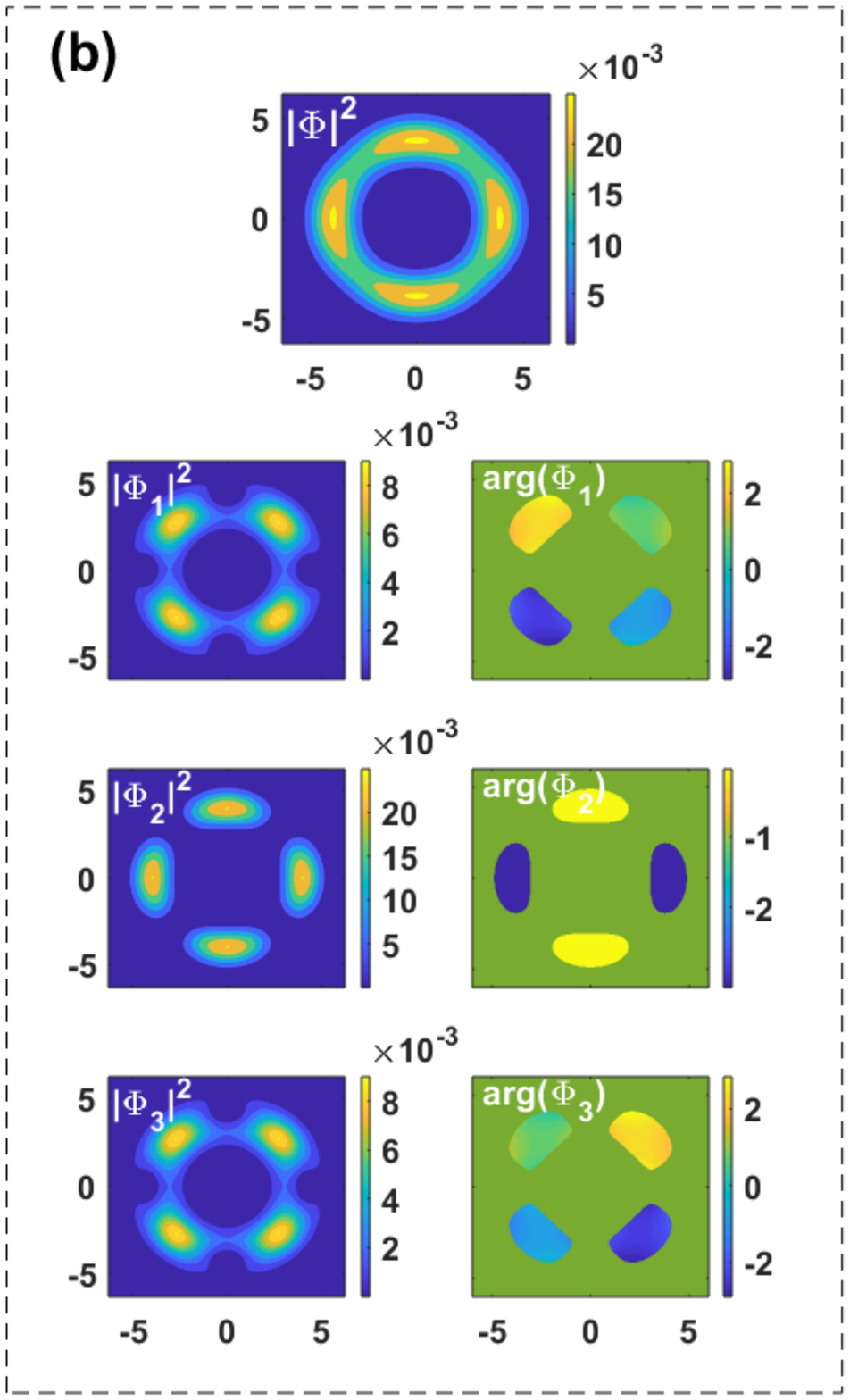}
	\includegraphics[width=0.24\textwidth,trim={110 40mm 30mm 80}, clip]{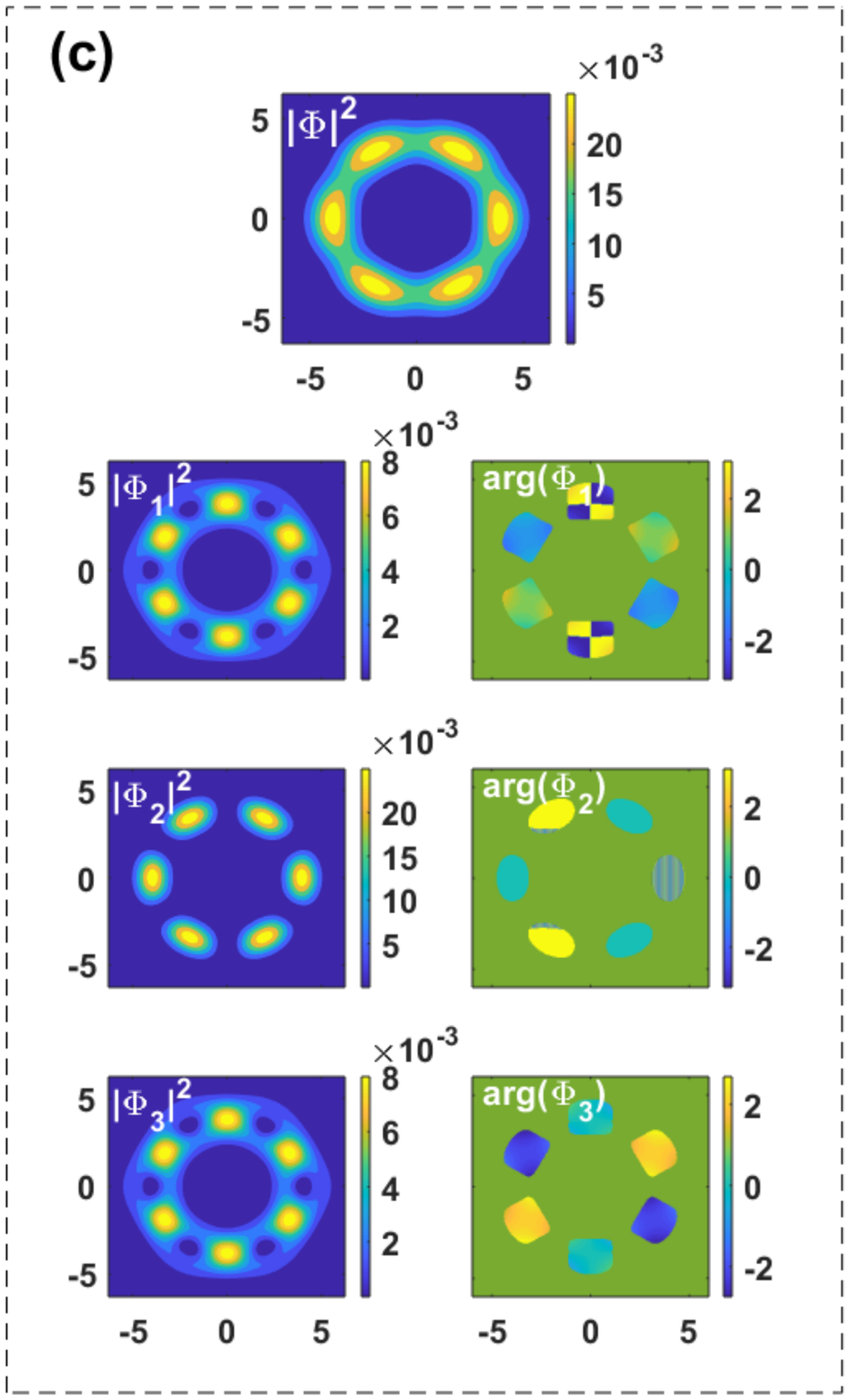}
	\includegraphics[width=0.24\textwidth,trim={110 40mm 30mm 80}, clip]{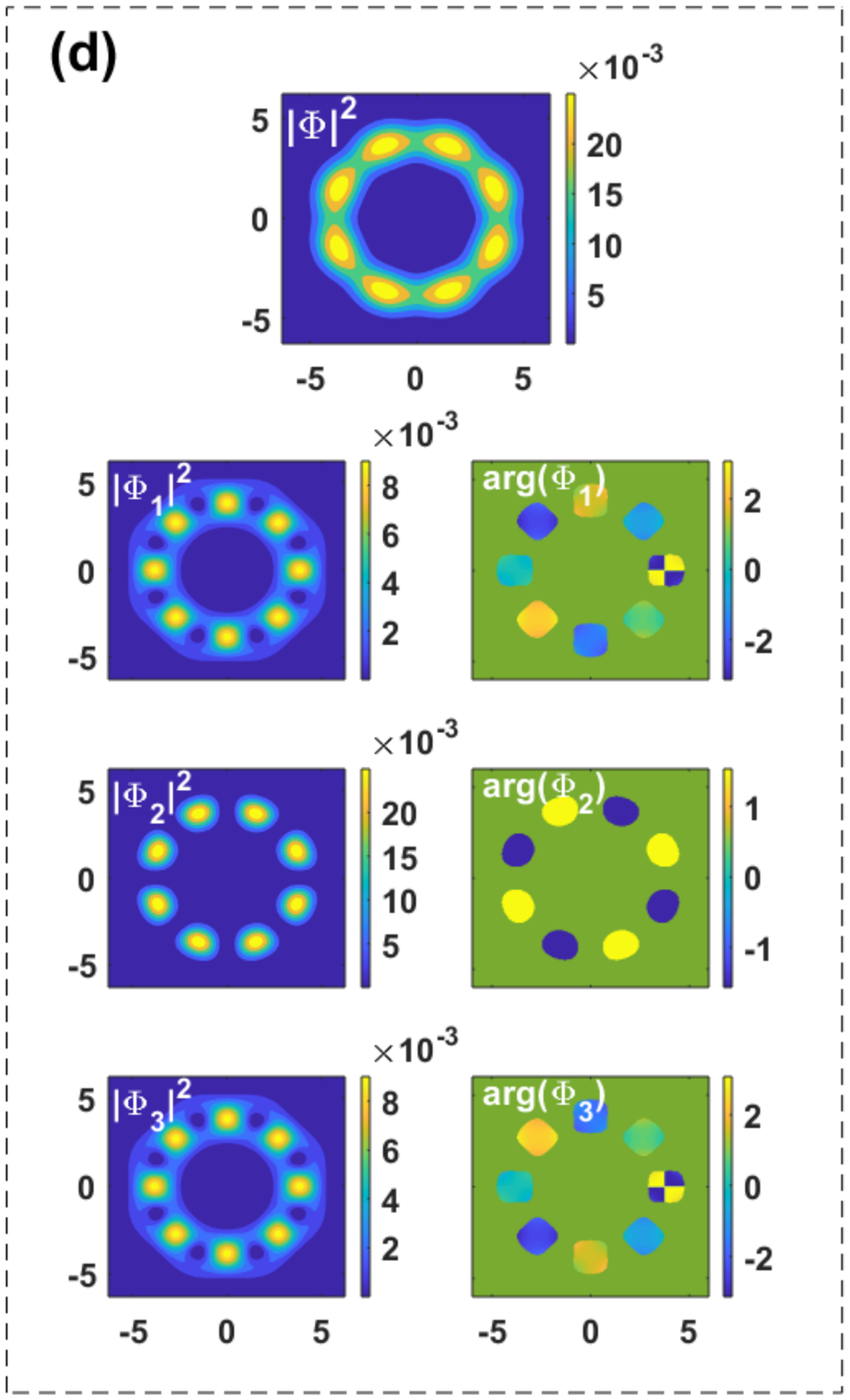}
	\caption{Ground states with the antiferromagnetic interaction $c_2=1$ for different spin-orbit coupling strength $\lambda$. (a) $\lambda=0.4$, (b) $\lambda=0.6$, (c) $\lambda=1.0$, (d) $\lambda=1.2$. Other parameters are $r_{0}=0.4$ and $c_0=10$. 
		In each block, the top panel is the total density distribution $|\Phi|^2=|\Phi_{1}|^2+|\Phi_{2}|^2+|\Phi_{3}|^2$, followed by density distribution of each component on the left and corresponding phase distribution of each component on the right.  	}
	\label{figone}
\end{figure*}

In this paper, we systematically characterize the ground-state phase diagram of a Rashba spin-orbit-coupled spin-1 BEC in a two dimensional toroidal trap. The phases are identified from both direct numerical calculations and an analytical study for a tight trap. When the toroidal trap is tight, the dynamics along radial direction can be frozen, two dimensional system is reduced to an one-dimension effective model only considering the dynamics along azimuthal direction. The effective model provides an analytical means to qualitatively understand  two dimensional numerical results.  A spin-1 spinor BEC features density-density interaction and spin-spin interaction with respective strength $c_0$ and $c_2$~\cite{Ueda2012}.   For an antiferromagnetic interaction $c_2>0$,  depending on the spin-orbit coupling strength, there are two phases:  persistent flows with the winding number $(-1,0,1)$ for three components and necklace states with even petal-number. For a ferromagnetic interaction $c_2<0$, the ground state is a persistent flow, and there is always a unit winding number difference between three components. We find that the origination and properties of all phases relate to extraordinary spin-flip symmetries. 

This paper is organized as follows. In Sec.~\ref{Phase}, we present the phase diagram from two dimensional numerical calculations. Features of the persistent flow and necklace state are addressed by a spin-flip symmetry. In Sec.~\ref{model}, we develop an  one-dimension analytical model to capture physics along the azimuthal direction.  From the model, all phases are identified using a variational wave function.  A clear physical picture is provided for the existence and properties of ground state. Especially we address why the necklace state must have even petal-number in a spin-1 BEC, while it would have odd number in a spin-1/2 analog.  Sec.~\ref{Conclusion} is the conclusion.

\section{Phase diagram}
\label{Phase}

The experimental realization of Raman-induced spin-orbit coupling in a spin-1 BEC~\cite{Spielman2016} stimulates to explore spin-orbit-coupled spinor BECs.  Ground states and collective excitations of a homogeneously spin-orbit-coupled spin-1 BEC have been investigated theoretically~\cite{Ohberg2014,Cole2015,Yu2016,Sun2016,Stringari2016}. With a toroidal trap, the system is described by the  Gross-Pitaevskii equation (GPE)~\cite{Yang2017, Sun2016},
\begin{equation}
\label{GP}
i\hbar \frac{\partial \Phi }{\partial t}=\left( H_\text{sin} + H_\text{int}  \right )\Phi.
\end{equation}
The spinor wave function $  \Phi=(\Phi_1,\Phi_2,\Phi_3)^T $ describes the occupation of three components. The single-particle Hamiltonian in Eq.~(\ref{GP}) is
\begin{equation}
\label{Single}
H_\text{sin}=\frac{p_{x}^2+p_{y}^2}{2m}+\lambda(F_{x}p_{y}-F_{y}p_{x})+V(r),
\end{equation}
with $m$ being the mass of the atom. $p_x$ and $ p_y$ are momenta along $x$ and $y$ directions respectively.  $(F_x,F_y,F_z)$ are spin-1 Pauli matrices. The Rashba spin-orbit coupling is $\lambda(F_{x}p_{y}-F_{y}p_{x})$ with the coupling strength $\lambda$.  The external trap is two dimensional toroidal, $V(r)=\frac{1}{2}m\omega_{r}^2(r-r_{0})^2$, here $r^2=x^2+y^2$, the radius of torus is  $r_{0}$, and $\omega_{r}$ is the trapping frequency. In the GPE, the nonlinear part is 
\begin{equation}
H_\text{int}=\left(\begin{array}{ccc} c_{0}\rho_{0}+c_{2}\rho_{z} & c_{2}\rho_{2} & 0 \\ c_{2}\rho_{2}^* & c_{0}\rho_{0} & c_{2}\rho_{2} \\ 0 & c_{2}\rho_{2}^* & c_{0}\rho_{0}-c_{2}\rho_{z} \end{array}\right),
\label{nonlinear}
\end{equation}	
where $\rho_{0}=|\Phi_{1}|^2+|\Phi_{2}|^2+|\Phi_{3}|^2$, 
$\rho_{z}=|\Phi_{1}|^2-|\Phi_{3}|^2$, and $\rho_{2}=\Phi_{2}^*\Phi_{1}+\Phi_{3}^*\Phi_{2}$.  The nonlinearity is characterized by the density-density interaction with the coefficient $c_0$ and spin-spin interaction with the coefficient $c_2$.  We numerically solve the GPE by the imaginary time evolution method to get ground states. In detail calculations, we use the dimensionless GPE by scaling the energy, length, time and $\lambda$  in units of $\hbar \omega_{r}$, $\sqrt{\hbar/m\omega_{r}}$, $1/\omega_{r}$ and $\sqrt{\hbar \omega_{r}/m}$ respectively. The wave function satisfies normalization condition, $\int dx dy ( |\Phi_{1}|^2+|\Phi_{2}|^2+|\Phi_{3}|^2 )=1$.

\begin{figure}[t]
	\includegraphics[ width=0.4\textwidth]{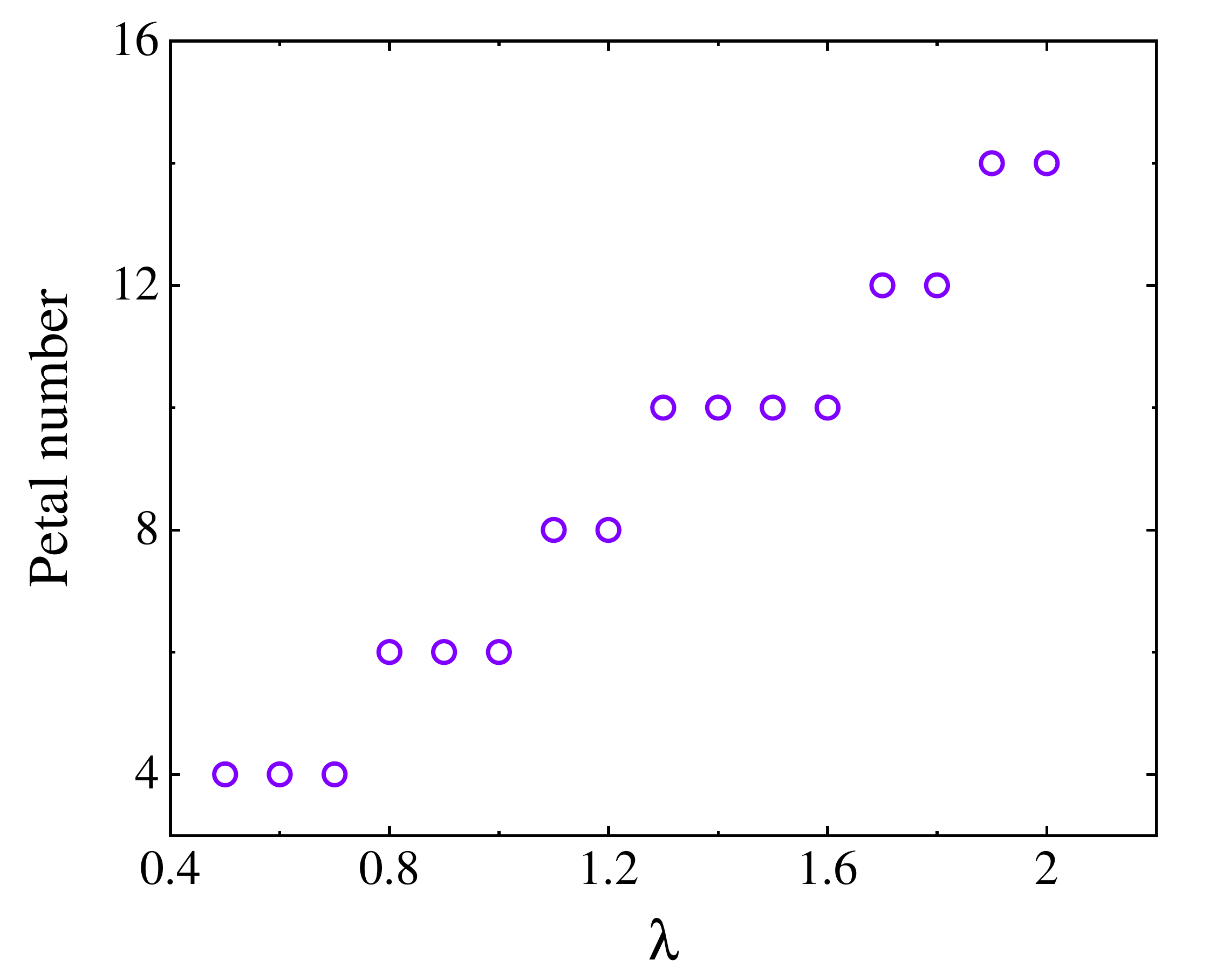}
	\caption{The number of petals in the necklace state as a function of the spin-orbit coupling strength $\lambda$. It is found that the petal-number is always even.  The other parameters are same as in Fig.~\ref{figone}. }
	\label{figtwo}
\end{figure}

Typical ground states for an antiferromagnetic interaction $c_2>0$ with different spin-orbit coupling strength $\lambda$ are demonstrated in Fig.~\ref{figone}. There are two different phases.  (1) When $\lambda$ is small, three components distribute homogeneously in the ring [see Fig.~\ref{figone}(a)].  The first and third components share the same density, $|\Phi_{1}|^2=|\Phi_{3}|^2$, and have an opposite-sign phase winding $\pm 1$. These two components support persistent flows with a unit phase winding.  While, there is no phase winding in the second component. The existence of phase winding makes the size of the first and third components is larger than that of the second component. (2) When $\lambda$ is relatively large, the ground state becomes necklacelike patterns, which are  shown in Fig.~\ref{figone}(b-d).  Three components have a same petal-number. The density of the first component is same as that of the third component, $|\Phi_{1}|^2=|\Phi_{3}|^2$.  The petals in the second component don't spatially overlap with these in  $\Phi_{1}$ and $\Phi_{3}$. The total density also shows a necklacelike geometry.   These states are reminiscent of stripe phases from a homogeneous spin-orbit-coupled BEC. In homogeneous system, Rashba spin-orbit coupling can induce density stripes whose orientation is spontaneously chosen~\cite{Zhai2011,Zhang2012}.  In the toroidal trap, the stripes are oriented along the azimuthal direction. Via this way, the boundaries of each petal in the necklace can be shortened for minimizing kinetic energy. The petal-number increases as a function of $\lambda$. In Fig.~\ref{figtwo}, we show the dependence of petal-number on $\lambda$, and find that the petal-number is always even and increases almost linearly with $\lambda$. The even-petal-number necklace state in spin-1 BECs is strikingly different from the only existed odd-number analog in spin-1/2~\cite{Busch2017}. It is noting that only necklace states with four times number of petals are numerically found in Ref.~\cite{Yang2017}.

These two different ground states have two common features.  From density and phase distributions in Fig.~\ref{figone}, we know that two ground states obey a same spin-flip symmetry $\hat{\mathcal{O}}$,
\begin{equation}
\hat{\mathcal{O}}=\mathcal{K} e^{i\pi F_y}= \mathcal{K} \begin{pmatrix} 0& 0&1 \\ 0 &-1 &0 \\ 1&0&0 \end{pmatrix}.
\label{symmetry}
\end{equation}
Here, $\mathcal{K}$ is the complex conjugate operator, and $e^{i\pi F_y}$ is the operator to rotate spins by the angle of  $\pi$ along the $F_y$ axis. The GPE and single-particle Hamiltonian $H_\text{sin}$ have the symmetry $\hat{\mathcal{O}}$. Ground states inherit the symmetry and are its eigenstates with eigenvalue $\pm 1$. This gives rise to $\Phi_1=\pm \Phi_3^*$ and $\Phi_2=\mp \Phi_2^*$. Therefore, the first and third components have a same density and the wave function of the second component is purely real or imaginary. The other common feature is the phase separation between $\Phi_2$ and $\Phi_1, \Phi_3$. The nonlinear part of the energy functional corresponding to the GPE is $E_\text{non}=E_\text{dd}+E_\text{ss}$. $E_\text{dd}=c_0/2 \int dxdy \left( |\Phi_1|^2 + |\Phi_2|^2 + |\Phi_3|^2  \right)^2$ is the density-density interaction energy, and 	the spin-spin interaction energy is $E_\text{ss}=c_2/2 \int dxdy [ (\Phi^\dagger F_x\Phi)^2+(\Phi^\dagger F_y\Phi)^2+(\Phi^\dagger F_z\Phi)^2 ]=c_2/2 \int dxdy [ ( |\Phi_1|^2 - |\Phi_3|^2)^2 +2 |\Phi_1|^2 |\Phi_2|^2+2  |\Phi_2|^2 |\Phi_3|^2  +\Phi_1\Phi_3\Phi_2^{*2}+ \Phi_1^*\Phi_3^*\Phi_2^{2}  ]$. Substituting the symmetry result $\Phi_1=\pm \Phi_3^*$ and $\Phi_2=\mp \Phi_2^*$ into the energy functional, we immediately realize that ground states having the symmetry  $\hat{\mathcal{O}}$  minimize the spin-spin interaction, i.e., $E_\text{ss}=0$. The density-density part becomes  $E_\text{dd}=c_0/2 \int dxdy ( 4 |\Phi_1|^4 + |\Phi_2|^4 + 4  |\Phi_1|^2 |\Phi_2|^2 )$, which is similar to a binary BEC. According to the phase separation condition of the binary BEC, $E_\text{dd}$ belongs to immiscible interactions, so $\Phi_1$ and $\Phi_2$ must be phase separated.  For persistent flow states [as shown in Fig.~\ref{figone}(a)], $\Phi_2$ and $\Phi_1, \Phi_3$ are spatially separated in the radial direction. While for necklace states, the separation is along the azimuthal direction.

\begin{figure*}[t]
	\includegraphics[ width=0.24\textwidth,trim={115 40mm 35mm 80}, clip]{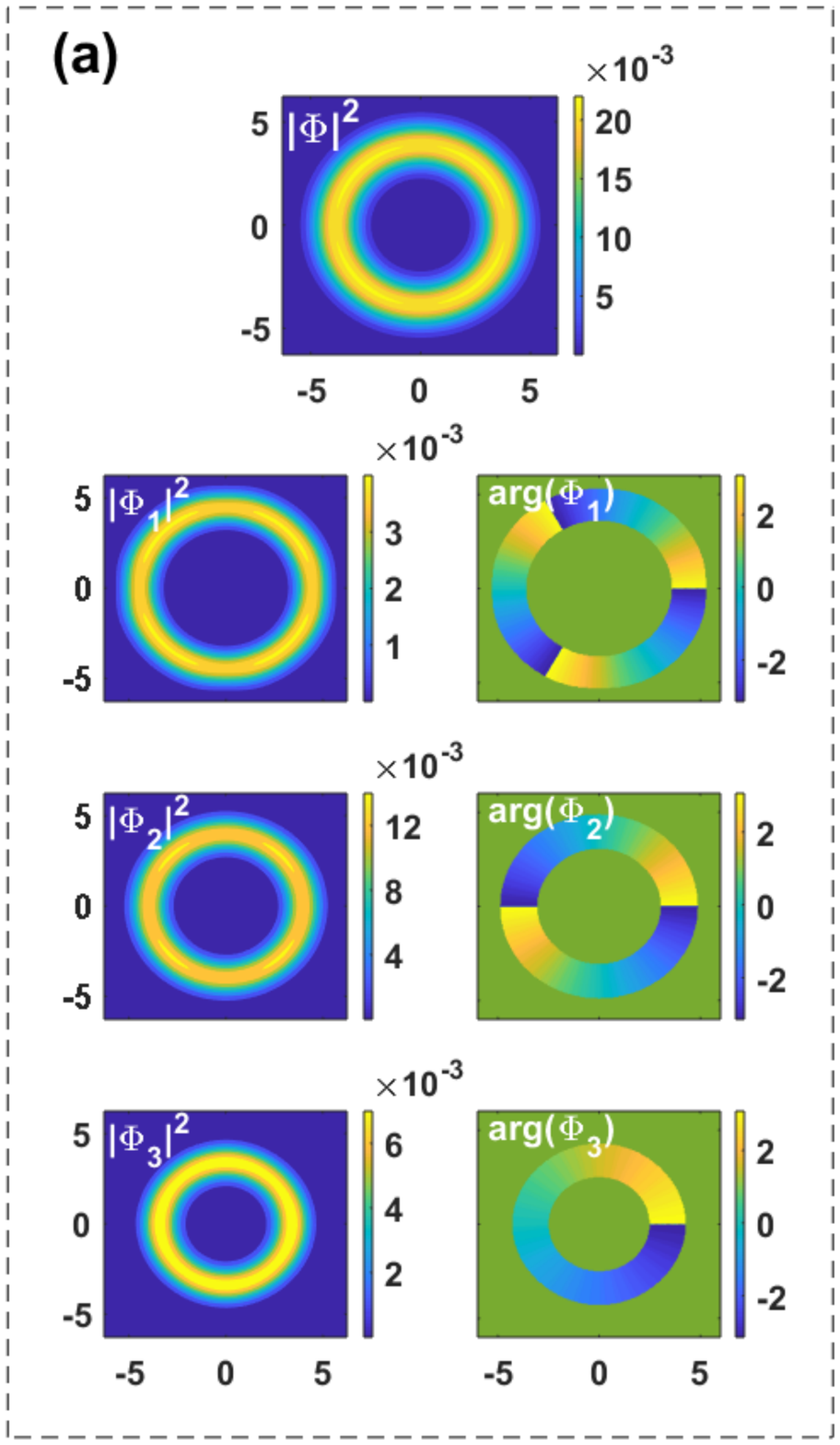}
	\includegraphics[ width=0.24\textwidth,trim={115 40mm 35mm 80}, clip]{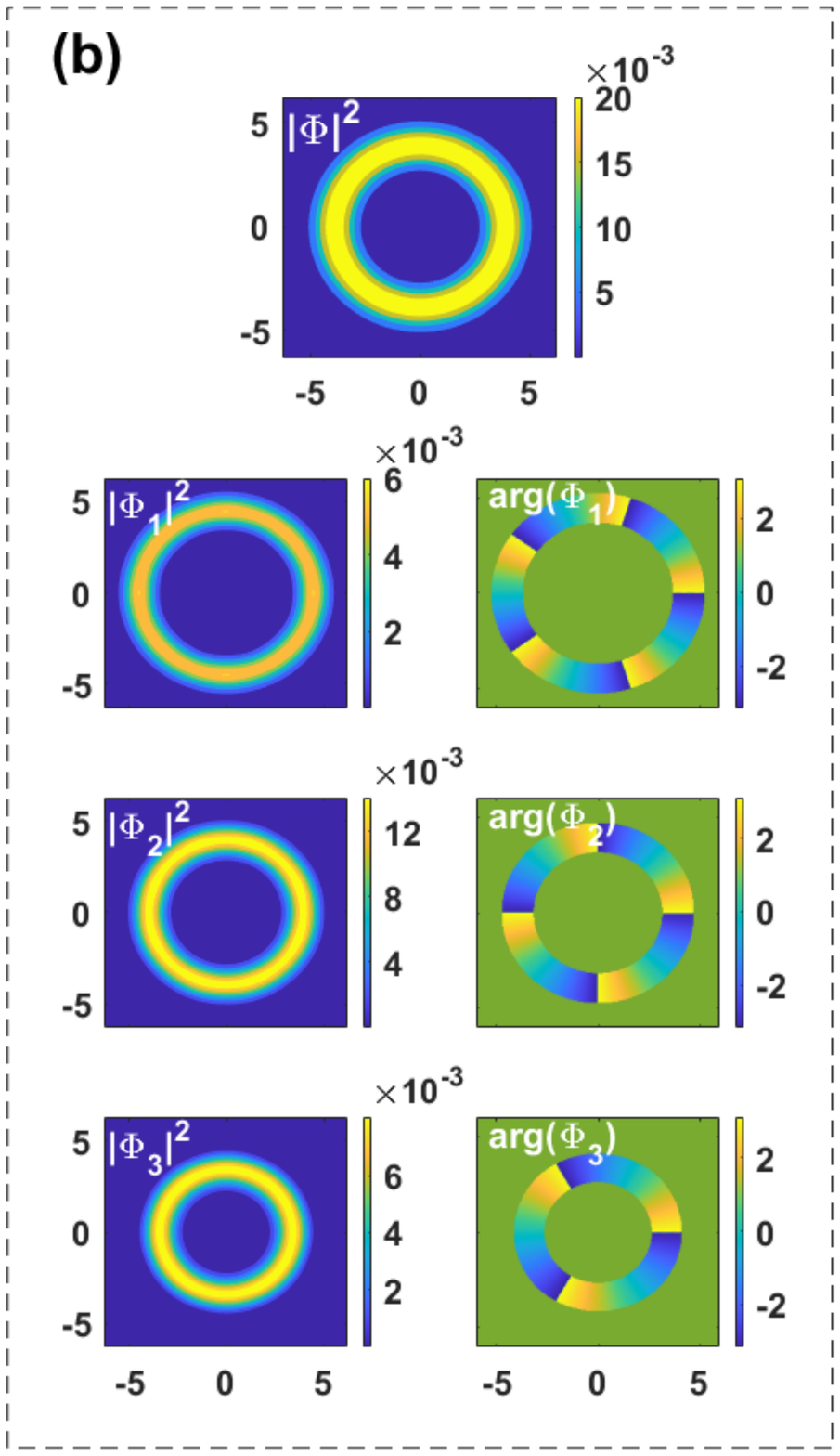}
	\includegraphics[ width=0.24\textwidth,trim={115 40mm 35mm 80}, clip]{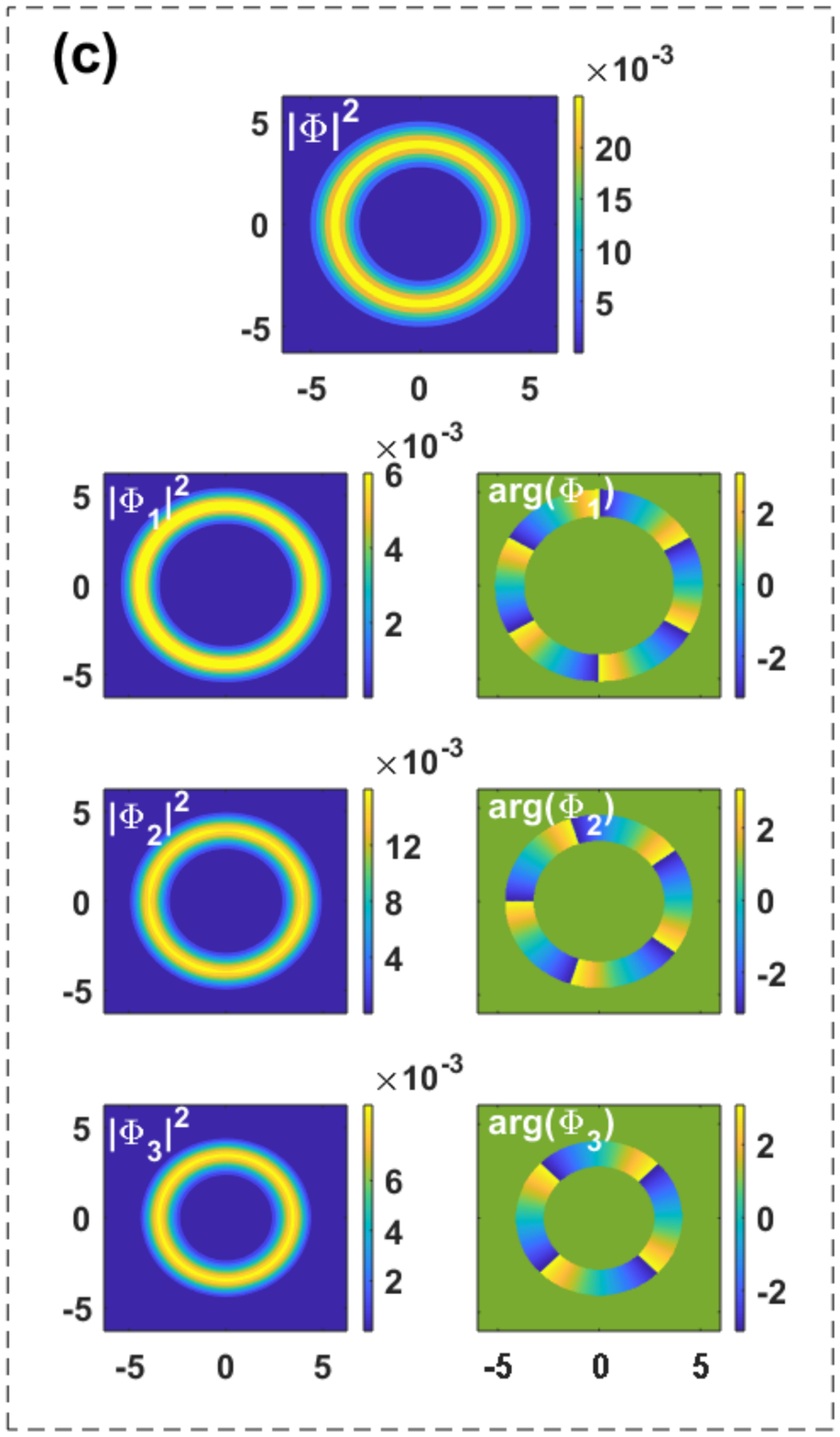}
	\includegraphics[ width=0.24\textwidth,trim={115 40mm 35mm 80}, clip]{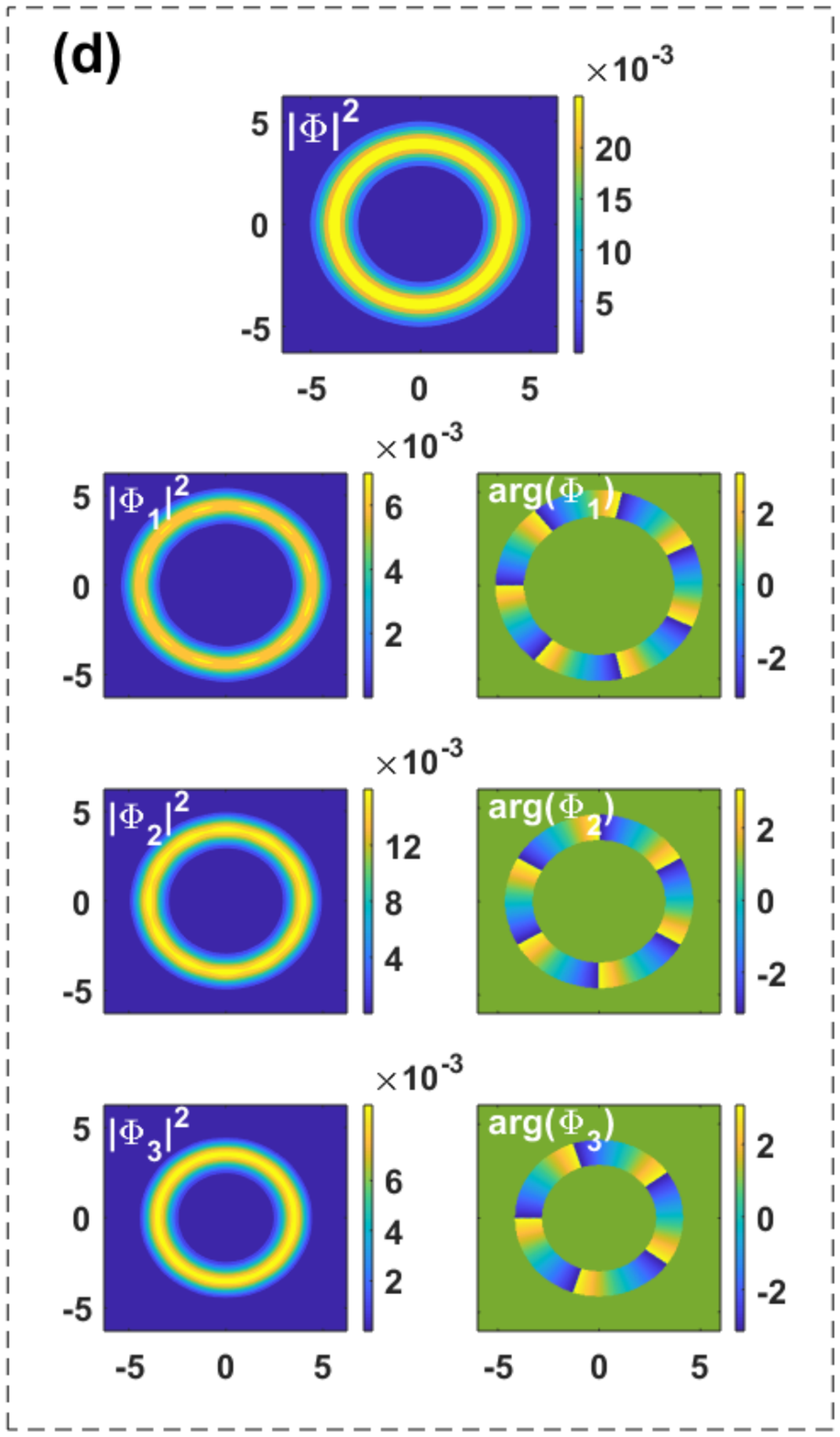}
	\caption{Ground states with the ferromagnetic interaction $c_2=-1$ for different spin-orbit coupling strength $\lambda$. (a) $\lambda=0.6$, (b) $\lambda=1.2$, (c) $\lambda=1.6$, (d) $\lambda=1.8$. Other parameters are $r_{0}=0.4$ and $c_0=10$. 
		In each block, the top panel is the total density distribution $|\Phi|^2=|\Phi_{1}|^2+|\Phi_{2}|^2+|\Phi_{3}|^2$, followed by density distribution of each component on the left and corresponding phase distribution of each component on the right.  }
	\label{figthree}
\end{figure*}

\begin{figure}[!]
	\includegraphics[ width=0.4\textwidth]{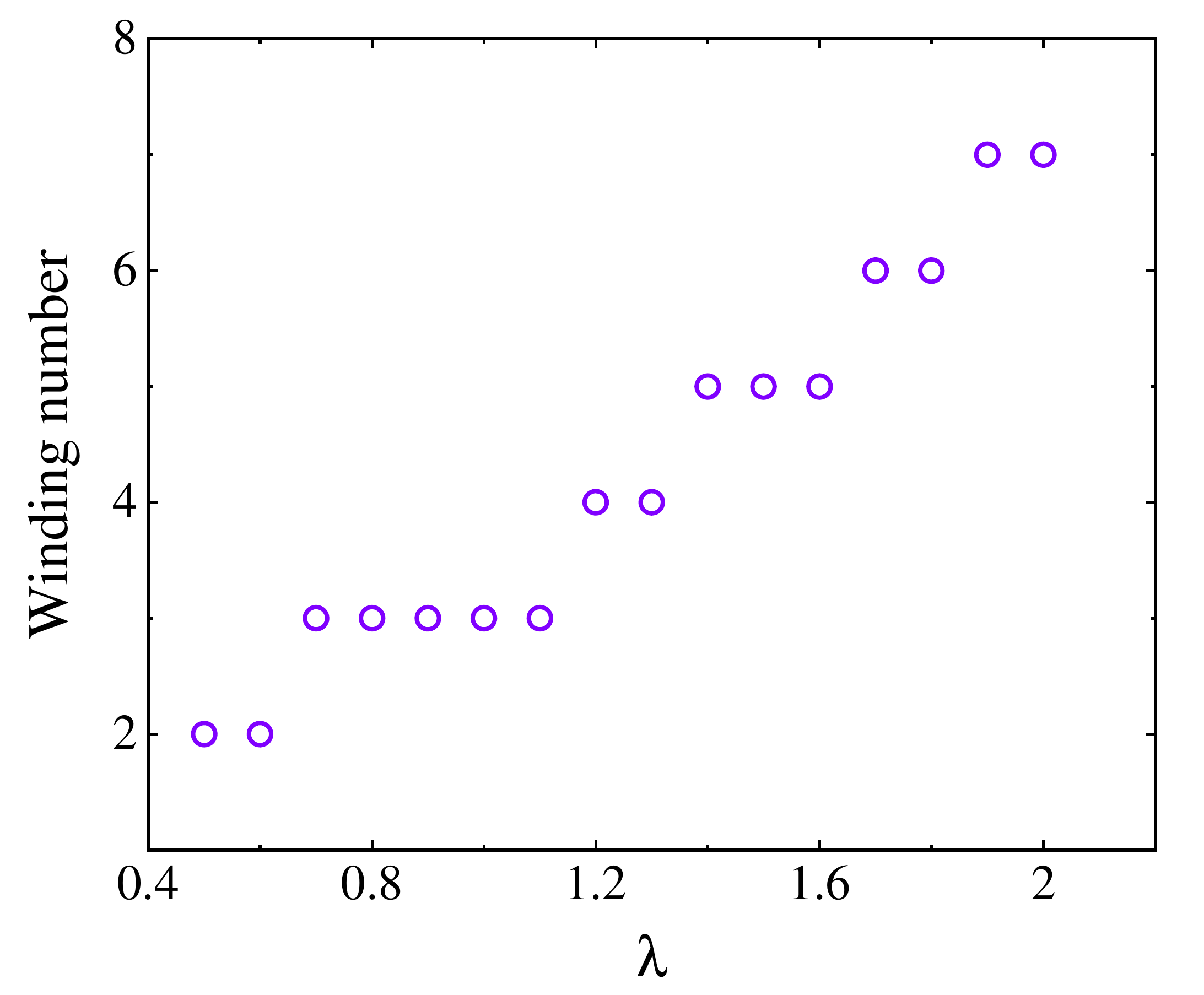}
	\caption{The winding number of the second component as a function of $\lambda$.  All parameters are same as in Fig.~\ref{figthree}.}
	\label{figfour}
\end{figure}

The above is the ground state with an antiferromagnetic interaction, we find that for a ferromagnetic interaction $c_2<0$, it becomes different. Typical results with $c_2=-1$ and different spin-orbit coupling strength $\lambda$ are depicted in Fig.~\ref{figthree}. The ground state always supports persistent flows and density distributes homogeneously along the radial direction. Every component carries nonzero phase winding.   The interesting is that there is always a unit phase winding difference between three components, which can be seen from the number of phase jumps in phase distributions in Fig.~\ref{figthree}.  The phase winding difference is also demonstrated from the size of density.  For fixed parameters, the persistent flow with a large number of phase winding has a larger density size. In Fig.~\ref{figthree}(a), for a small $\lambda$, the winding number for $\Phi_1$, $\Phi_2$  and $\Phi_3$ are $-3$,  $-2$, and $-1$ respectively, and the density size decreases. Increasing the spin-orbit coupling strength, the phase winding in each component increases as shown in Fig.~\ref{figthree}(b-d).
The dependence of the winding number in the second component on $\lambda$ is demonstrated in Fig.~\ref{figfour}.
The winding number increases almost linearly as a function of 
$\lambda$.

Two dimensional numerical phase diagram shows interesting characteristics. According to antiferromagnetic or ferromagnetic interactions, two different persistent flow states exist, and necklace states have even petal-number.  In the following, we develop an one-dimension effective model to provide analytical insights into the existence and unique properties of these states.

\section{EFFECTIVE MODEL}
\label{model}

All states in the previous reflect that the azimuthal effect plays an important role. We assume the toroidal trap is very tight, i.e., $\omega_r$ is a large scale. In this condition, the dynamics along the radial direction is frozen into the ground state of the harmonic trap. After integrating over the radial degree of freedom, the dynamics along the azimuthal direction would shine through~\cite{Zulicke2012, Busch2017, Klapwijk2002}. The single-particle Hamiltonian in Eq.~(\ref{Single}) reduces to one dimensional $H_\text{eff}$ which effectively describes the azimuthal effect~\cite{Klapwijk2002},
\begin{equation}
\begin{aligned}
H_\text{eff}=&\left( -i\frac{\partial}{\partial\phi} \right)^2+\bar{\lambda} \left(\cos{\phi}F_{x}+\sin{\phi}F_{y} \right )\left(-i\frac{\partial}{\partial \phi}\right)\\
&-i\frac{\bar{\lambda}}{2}\left(\cos{\phi}F_{y}-\sin{\phi}F_{x}\right).
\end{aligned}
\label{effective}
\end{equation}
Here, $\phi$ is the azimuthal coordinate. $H_\text{eff}$ is dimensionless, we use the unit of energy as $\hbar^2/(2mr_0^2)$, and $\bar{\lambda}=2mr_0\lambda/\hbar$.
After considering the nonlinear part in  Eq.~(\ref{nonlinear}), the total energy becomes,
\begin{align}
E_\text{tot}&= \frac{1}{2\pi}\int_0^{2\pi}d\phi \left( \bar{\Phi}^\dagger  H_\text{eff}  \bar{\Phi}  \right)+ \frac{\bar{c}_0}{4\pi} \int_0^{2\pi}d\phi |  \bar{\Phi} |^4   \notag \\
&  +  \frac{\bar{c}_2}{4\pi}  \int_0^{2\pi}d\phi  \left[ \left(\bar{\Phi}^\dagger F_x\bar{\Phi}  \right)^2+ \left(\bar{\Phi}^\dagger F_y\bar{\Phi}  \right)^2+ \left(\bar{\Phi}^\dagger F_z\bar{\Phi} \right)^2 \right].
\label{totalenergy}
\end{align}
Here,  $ \bar{c}_0= 2mr_0^2/\hbar^2 \sqrt{ m\omega_r/(2\pi \hbar ) } c_0  $ and $ \bar{c}_2= 2mr_0^2/\hbar^2 \sqrt{ m\omega_r/(2\pi \hbar ) } c_2  $. The reduced wave function is $ \bar{\Phi} (\phi)=( \bar{\Phi}_1(\phi),  \bar{\Phi}_2(\phi),  \bar{\Phi}_3(\phi))^T	$, and $ |\bar{\Phi}|^2=   |\bar{\Phi}_1|^2   + |\bar{\Phi}_2|^2 + |\bar{\Phi}_3|^2  $.

The effective Hamiltonian $H_\text{eff}$ conserves $J_z$ with its definition being
\begin{equation}
J_z=-i \frac{\partial }{ \partial \phi} +F_z,
\end{equation}
i.e., $[J_z, H_\text{eff}]=0 $. The conservation makes $H_\text{eff}$ invariant under a rotation $U=\exp(i\theta J_z)$, where $\theta$ is an arbitrary angle. Therefore, $H_\text{eff}$ and $U$ have same eigenstates, which can be constructed from  $U$ as,
\begin{equation}
\bar{\Phi}=e^{in\phi}\left(\begin{array}{ccc} e^{-i\phi}\Phi'_{1} \\ \Phi'_{2} \\  e^{i\phi}\Phi'_{3} \end{array}\right),
\end{equation}
where  $ \Phi'_{1}, \Phi'_{2}$ and $\Phi'_{3} $ are independent on $\phi$. $n$ is an integer number and characterizes phase winding. Three components carry phase winding $(n-1, n, n+1)$.  There is a unit phase winding difference between them. With the help of $ \bar{\Phi}$, the effective Hamiltonian $H_\text{eff}$ can be diagonalized to get eigenvalues $E(n)$, from which we immediately realize that $E(n)=E(-n)$. Physically, the degeneracy of $E(n)$ and $E(-n)$ originates from the symmetry $\hat{\mathcal{O}}$ defined in Eq.~(\ref{symmetry}). Therefore, the lowest energy state of the single-particle effective Hamiltonian $H_\text{eff}$ is twofold degenerate.

The twofold degenerate lowest energy state offers an interesting accommodation for atoms to condense into. In the presence of the mean-field interactions, we construct the ground state ansatz by considering the superposition of the lowest energy state~\cite{Stringari2012, Sun2016},
\begin{equation}
\bar{\Phi}^\text{(gs)}=C_{+}\bar{\Phi}_{n} + C_{-} \hat{ \mathcal{O}} \bar{\Phi}_{n}.
\end{equation}
Here,
\begin{equation}
\bar{\Phi}_{n}=
e^{in\phi}\left(\begin{array}{ccc} e^{-i\phi}\cos{\alpha}\cos{\beta} \\ -\sin{\alpha} \\  e^{i\phi} \cos{\alpha}\sin{\beta}\end{array}\right),
\label{eigen}
\end{equation}
which is an eigenstate of $H_\text{eff}$.  The superposition coefficients $C_{+}$ and $C_{-}$ satisfy $|C_{+}|^2+|C_{-}|^2=1$.  Substituting the trial wave function $\bar{\Phi}^\text{(gs)}$ into the total energy $E_\text{tot}$ in Eq.~(\ref{totalenergy}), we get $E_\text{tot}(n, \alpha, \beta, C_{+},C_{-})$.  All parameters $n, \alpha, \beta, C_{+}$ and $C_{-}$, determining properties of the ground state, shall be fixed by minimizing the total energy $E_\text{tot}$. 

According to the resulted parameters from the minimization,  the ground state has the following phases:

(1) For antiferromagnetic interactions $\bar{c}_2>0$, there are two phases. We take $\bar{c}_0=10$ and $\bar{c}_2=1$ as an example. (I) When  $\bar{\lambda} \le 0.86$, the results are $n=0$, $\beta=3\pi/4$, and one of the $C_{+}$ and $C_{-}$ is zero, i.e.,  $C_{+}=0$, $C_{-}=1$, or $C_{+}=1$, $C_{-}=0$. With these parameters, it can be seen that $\hat{\mathcal{O}} \bar{\Phi}_{n=0} = - \bar{\Phi}_{n=0} $.  $\bar{\Phi}_{n=0}$ and $\hat{\mathcal{O}} \bar{\Phi}_{n=0} $ are a same state up to a sign difference. The ground state supports persistent flows. The phase winding of three components is always $-1$, $0$ and $1$. $|\bar{\Phi}_1^\text{(gs)}|^2=|\bar{\Phi}_3^\text{(gs)}|^2$, the size of which is always larger than that of $|\bar{\Phi}_2^\text{(gs)}|^2$.  This phase corresponds to the two dimensional analog as shown in Fig.~\ref{figone}(a). (II) When  $\bar{\lambda} > 0.86$, results of the minimization are  $n \ne0$ and $C_{+}=1/\sqrt{2}$, $ C_{-}=\pm1/\sqrt{2}$. The ground state is the equal superposition of $\bar{\Phi}_{n}$ and $\hat{ \mathcal{O}} \bar{\Phi}_{n}$, and thus is an eigenstate of $\hat{ \mathcal{O}} $. Because of the superposition, the density of the ground state is periodically modulated along the azimuthal direction,  $|\bar{\Phi}_1^\text{(gs)}|^2= |\bar{\Phi}_3^\text{(gs)}|^2 = \cos^2\alpha \left[ 1\pm \sin(2\beta )\cos(2n\phi) \right] $, $|\bar{\Phi}_2^\text{(gs)}|^2= 2\sin^2\alpha \left[ 1-\cos(2n\phi)\right]$, here $\pm$ depends on the sign of $C_{-}$. Such density modulated ground states correspond to necklace states found in Fig.~\ref{figone}(b-d). The nature of necklace states is that the period of three components is same and is an even number $2n$.  

(2)	For ferromagnetic interactions $\bar{c}_2<0$, the minimization always chooses  $C_{+}=0$, $C_{-}=1$ or $C_{+}=1$, $C_{-}=0$. The ground state is a persistent flow with phase winding number $(n-1,n,n+1)$ or $(-n-1,-n,-n+1)$ for three components. These two configurations are spontaneously chosen. The unique feature of such a ground state is that there is a unit phase winding difference between three components. These states correspond to two-dimensional persistent flows found in Fig.~\ref{figthree}, where only the configuration $(-n-1,-n,-n+1)$ is shown.

The possible existence of the even-petal-number necklace state in spin-1 BECs is due to the cooperation of symmetries  $\hat{ \mathcal{O}}$ and $J_z$. It is interesting to compare with a spin-1/2 system, where only the odd-petal-number state can exist. The effective azimuthal Hamiltonian for a Rashba coupled spin-1/2 is $H'_\text{eff}= (-i\partial/\partial \phi)^2+ \bar{\lambda} ( \cos\phi \sigma_x +\sin\phi \sigma_y) (-i\partial/\partial \phi) -i \bar{\lambda}/2 ( \cos\phi \sigma_y -\sin\phi \sigma_x  ) $, which is similar as  Eq.~(\ref{effective}), but replacing spin-1 matrices $F$ by spin-1/2 Pauli matrices $\sigma$~\cite{Zulicke2012, Busch2017}. The conservation of $J'_z=-i\partial / \partial \phi +\sigma_z/2$ for $H'_\text{eff}$ requires its eigenstates as $ \bar{\Phi}'_n=e^{in\phi} ( \Phi'_1, e^{i\phi}\Phi'_2  )^T $. The symmetry $\hat{ \mathcal{O}}'=\mathcal{K} e^{i\pi \sigma_y/2} $ gives rise to the degeneracy of  $\bar{\Phi}'_n$ and $\hat{ \mathcal{O}}' \bar{\Phi}'_n$. Therefore, the possible ground state is the superposition of $\bar{\Phi}'_n$ and $\hat{ \mathcal{O}}' \bar{\Phi}'_n$ with the corresponding density $\propto  \cos[(2n+1)\phi]$. The necklace state in the spin-1/2 system takes odd petal-number. It is noting that there is a 1/2 in the symmetries $\hat{ \mathcal{O}}'$ and  $J'_z$, which is due to the $SU(2)$ nature of spin-1/2 spins. This unique $SU(2)$ property makes petal-number in necklace states of a spin-1/2 BEC odd. 

\begin{figure}[t!]
	\includegraphics[ width=0.4\textwidth]{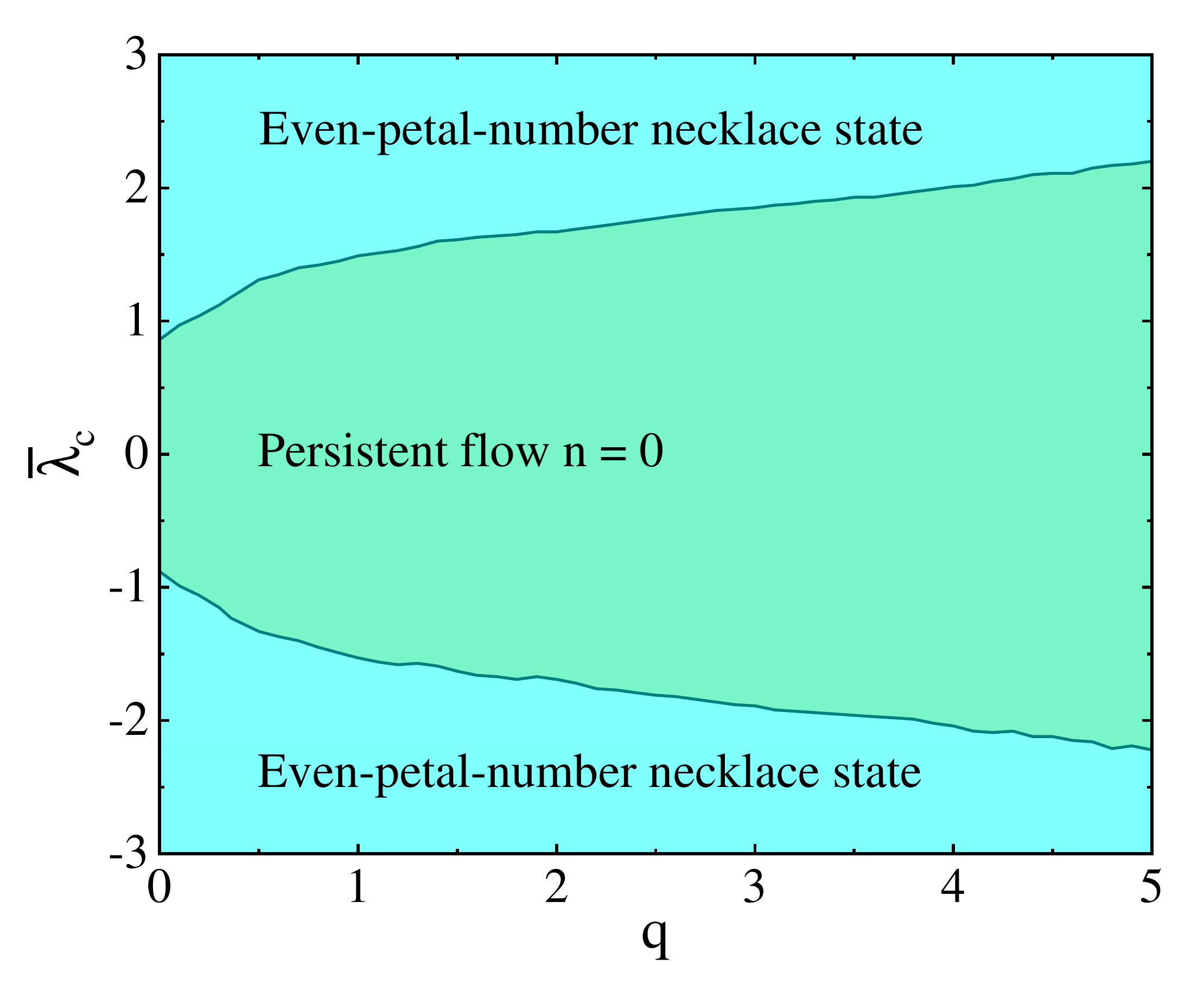}
	\caption{The critical value of the spin-orbit coupling strength $\bar{\lambda}_{c}$ between the persistent flow and necklace state as a function of the quadratic Zeeman effect $q$ for an antiferromagnetic interaction $\bar{c}_2=1$ from the one dimensional effective model. The other parameter is $\bar{c}_2=10$.}
	\label{figfive}
\end{figure} 

We conclude that it is  the symmetries  $\hat{\mathcal{O}}$ and $J_z$ of the Rashba spin-orbit coupling that leads to the existence of the persistent flow and necklace state. In spin-1 BECs, there is a unit phase winding difference between three components for persistent flows, and necklace states always have even petal-number.  

At last, we would like to address the effect of the quadratic Zeeman coupling on above results. The quadratic Zeeman coupling plays an important role in spin-1 spinor BECs~\cite{Ueda2012}.  It is $qF_z^2$, which should be incorporated into $H_\text{sin}$ in Eq.~(\ref{Single}) and $H_\text{eff}$ in Eq.~(\ref{effective}). Here, $q$ describes the strength of the quadratic Zeeman term.  In the Raman-induced spin-orbit-coupled spin-1 BEC experiment, it can be tuned~\cite{Spielman2016}. It is interesting to find that the quadratic Zeeman term doesn't destroy the symmetries  $\hat{\mathcal{O}}$ and $J_z$. Therefore, its existence doesn't quantitatively change the results from two-dimension numerical calculations and from the one-dimension analytical model. We find that it just slightly modifies the demarcation between the persistent flow and necklace state for antiferromagnetic interactions. In Fig.~\ref{figfive}, we show the dependence of the critical spin-orbit coupling strength $\bar{\lambda}_c$ on the quadratic Zeeman effect $q$.  For antiferromagnetic interactions,  when $|\bar{\lambda}| < |\bar{\lambda}_c|$, the ground state is a persistent flow with $n=0$ [see Eq.~(\ref{eigen})], when  $|\bar{\lambda}| > |\bar{\lambda}_c|$, the ground state is an even-petal-number necklace state. There is a slight increase of $|\bar{\lambda}_c|$ as a function of $q$.  We also note that in Fig.~\ref{figfive} there is a symmetry between $\bar{\lambda}>0$ and  $\bar{\lambda}<0$, this is because that the effective Hamiltonian $H_\text{eff}+qF_z^2 $ has a spin rotation symmetry $ e^{i\pi F_z}$, the physics in the $\bar{\lambda}<0$ regime is the same as that in the regime of $\bar{\lambda}>0$.

\section{conclusion}
\label{Conclusion}

We systematically study the ground-state phase diagram in a Rashba spin-orbit-coupled spin-1 BEC trapped in a two dimensional toroidal trap. The spin-flip symmetries of the spin-orbit coupling endow new features to the persistent flow and support the even-petal-number necklace state. The chosen phases depend on the sign of  spinor's spin-spin interaction. The toroidal trapped spin-orbit-coupled BEC provides an experimentally accessible playground to investigate necklace states. They emerge as ground states  with spontaneous breaking of the continuous rotation symmetry. Therefore, they are always dynamically unstable, which is in favor of experimental observation. The petal-number of the necklace state can be tuned by changing the spin-orbit coupling strength. The number is always odd in spin-1/2 BECs, while it is even in spin-1 systems. 

\begin{acknowledgments}
	
	This work is supported by National Natural Science Foundation of China with Grants Nos.~11974235 and 11774219.
	
\end{acknowledgments}


\end{document}